\documentclass[]{aa}

\usepackage{txfonts,color, graphicx,rotating, natbib}

 \usepackage[pdftex,
        colorlinks=false,
        urlcolor=blue,       
        filecolor=green,     
        linkcolor=red,       
        pdftitle={Paper_dwarf},
        pdfauthor={Zhukovska},
        pdfsubject={article},
        pdfproducer={pdfLaTeX},
        pdfpagemode=None,
        bookmarksopen=true]{hyperref}
\begin{document}
\def\HI{\ion{H}{i} }
\newcommand{\OH}{\ensuremath{ \rm 12+log(O/H)}}
\def\H2{\ensuremath{ \rm H_{\rm 2} }}
\newcommand{\mum}{\ensuremath{\rm \, \mu m}}
\newcommand{\Ms}{\ensuremath{\rm \,M_{\sun}}}
\newcommand{\Zs}{\ensuremath{\rm \,Z_{\sun}}}
\newcommand{\Mspc}{\ensuremath{\rm \,M_{\sun}\,pc^{-2}}}
\newcommand{\Msyr}{\ensuremath{\rm \,M_{\sun}\,yr^{-1}}}
\newcommand{\pyr}{\ensuremath{\rm \,yr^{-1}}}
\newcommand{\cmc}{\ensuremath{\rm \,cm^{-3}}}
\newcommand{\kms}{\ensuremath{\rm \,km\,s^{-1}}}
\newcommand{\ddt}[1]{{{\rm d}\, {#1} \over{\rm d}\,t}}
\newcommand{\red}[1]{{\color{red} #1}}
\newcommand{\blue}[1]{{\color{blue} #1}}

\author{Svitlana Zhukovska\inst{1,}\inst{2}}
\institute{Max Planck Institute for Astronomy, K\"onigstuhl 17, D-69117 Heidelberg, Germany
\and
Zentrum für Astronomie der Universit\"at Heidelberg, Institut f\"ur Theoretische Astrophysik, Albert-Ueberle-Str. 2, 69120 Heidelberg, Germany}
\title{Dust origin in late-type dwarf galaxies:\\ ISM growth vs. type II supernovae}
\titlerunning{Dust origin in late-type dwarf galaxies}
\abstract
{
}
{
We re-evaluate the roles of different dust sources in dust production as a function of metallicity in late-type dwarf galaxies, with the goal of understanding the relation between dust content and metallicity.
}
{
The dust content of late-type dwarf galaxies with episodic star formation is studied with a multicomponent model of dust evolution, which includes dust input from AGB stars, type II SNe, and dust growth by accretion of atoms in the ISM.}
{
Dust growth in the ISM becomes an important dust source in dwarf galaxies on the timescale of $10^8$ to a few $10^9$~yr. It increases the dust-to-gas mass ratio (DGR) during post-burst evolution, unlike type II SNe, which eject grains into the ISM only during starbursts. Before the dust growth in the ISM overtakes the dust production, AGB stars can be major sources of dust in  metal-poor dwarf galaxies.
Our models reproduce the relation between the DGR and oxygen abundance, derived from observations of a large sample of dwarf galaxies. The steep decrease in the DGR at low metallicities is explained by the relatively low efficiency of dust condensation in stars. The scatter observed at higher metallicities is determined mainly by different metallicities for the transition from stardust- to ISM-growth-dominated dust production, depending on the star formation history. 
In galaxies with episodic star formation, additional dispersion in the DGR is introduced by grain destruction during starbursts, followed by an increase in the dust mass owing to dust growth in the ISM during post-burst evolution. 
We find that the carbon-to-silicate ratio changes dramatically when the ISM growth becomes the dominant dust source, therefore this ratio can be used as an indicator of the transition.
} 
{
The observed relation of the DGR versus metallicity in dwarf galaxies favours low-condensation efficiencies in type II SNe, together with an increase in the total dust mass by means of dust growth in the ISM.
}

\date{}
\maketitle

\section{Introduction}
The origin of interstellar grains has been highly debated. Grains are formed in the stellar winds of evolved stars and expanding ejecta of supernovae, which have been considered the major sites of dust formation. However, high depletions of gas-phase element abundances in dense gas cannot be reproduced  only with the stellar dust production. They require additional dust growth by accretion of gas-phase species on grain surface directly in the ISM \citep[e.g.,][]{Dwek:1980p490, Weingartner:1999p6573, Jenkins:2009p2144}. Indeed, dust growth in the ISM is considered the dominant dust source in the Milky Way \citep{Dwek:1998p67, Zhukovska:2008bw, Draine:2009p6616}. The low values of element depletions in diffuse gas indicate efficient dust destruction by supernova shocks \citep{Draine:1979p1036, Barlow:1978vx, Dwek:1980p490, Jones:1994p1037, Jones:1996p6593, 2011A&A...530A..44J}.  

Details of the dust growth in the ISM are poorly understood (\citealt{Draine:2009p6616}; but see \citealt{Krasnokutski:2013wz}). It clearly depends on the collision rates of dust-forming gas species with grain surfaces, hence it strongly depends on the gas-phase element abundances, and thus on the metallicity. In the Milky Way, the critical metallicity for the efficient dust growth by accretion is $Z \sim 0.001$ for silicate and somewhat higher for carbonaceous and iron grains, 0.004 and 0.014, respectively \citep{Zhukovska:2008p7215, Zhukovska:2009p7232}. Recently, \cite{Asano:2013kl} have studied the critical role of metallicity for the dust growth in galaxies with the continuous star formation rate. They find that it depends on the star formation timescales, $Z_{\rm crit} \sim \tau_{\rm SF}^{-1/2}$. 

In the present paper, we investigate the transition from stardust-dominated to ISM  growth-dominated dust production in late-type dwarf galaxies, blue compact dwarf (BCD), and dwarf irregulars. They are gas-rich, relatively unevolved systems, with the metallicities varying over a wide range down to the lowest values known in the local Universe \citep[$Z \sim 1/40\ \Zs$ in BCD SBS 0335-052, ][]{Izotov:1997ip}. The upper range of metallicities in BCDs, $\OH \sim 8.4$ \citep{Izotov:1999ch}, slightly exceeds the metallicity of the Large Magellanic Cloud, where  the dust growth in the ISM is expected to be important \citep{Matsuura:2009p363, Zhukovska:2013vg}. Star formation in dwarf galaxies occurs in bursts separated by quiescent periods \citep[e.g.,][]{Recchi:2002p6651, Tolstoy:2009cl, Lanfranchi:2003p1954, Yin:2011he, Bradamante:1998to}, therefore they permit studying the effect of fast chemical enrichment during starbursts and slow post-burst galactic evolution on its dust content.  

Despite the low metallicities, late-type dwarfs indicate the presence of substantial amounts of dust characterised by large dispersion in the dust-to-gas ratio (DGR) \citep[e.g.,][]{Lisenfeld:1998p491}. The dust-to-gas ratio of a sample of galaxies plotted as a function of oxygen abundances provides an insight into the relation between dust-to-gas ratio and metallicity. This relation is crucial for detailed numerical hydrodynamical simulations of the ISM evolution, for which dust abundance is usually taken as a fixed fraction of metals. This assumption can be justified for spiral galaxies with solar and supersolar metallicities, but not for metal-poor dwarf galaxies \citep[e.g.,][]{Dwek:1998p67, Lisenfeld:1998p491, Draine:2007de, Engelbracht:2008p4082, Galliano:2005p6620, Galliano:2011gy}. \cite{Engelbracht:2008p4082} find that the relation between the DGR and oxygen abundance  \OH\ in dwarf galaxies flattens out at low metallicities $\OH < 8$. However, more recent studies have shown that it actually steepens at low metallicities \citep{HerreraCamus:2012fb, Galametz:2011kxa, RemyRuyer:2013uu}. We will investigate how these variations in gas-to-dust ratio can be explained with the evolutionary dust models.

Dust in dwarf galaxies has been studied with the dust and gas chemical evolution models in the literature \citep{Lisenfeld:1998p491, Hirashita:1999p2082, Hirashita:2002p6608, Hirashita:2002p2081, Calura:2008p1752}. These models ascribe the origin of interstellar dust to the SN type II \citep{Lisenfeld:1998p491, Hirashita:2002p6608}, which require quite a high degree of condensation in dust in SN ejecta, about 10\% \citep{Hirashita:1999p2082, Hirashita:2002p2081}. This value is significantly higher than estimates from various recent observational and theoretical studies (\citealt{Bianchi:2007p2222, Ercolano:2007p2083, Zhukovska:2008bw,  Kozasa:2009p1041, Cherchneff:2009p3933, Cherchneff:2010p6001, Fox:2010gi}; but see  \citealt{Matsuura:2011ij, Gomez:2012fm, Lakicevic:2011fl}). Therefore, dust growth by accretion in the ISM may be more important in dwarf galaxies than previously thought.

Our main goal is to investigate the relation between dust content and metallicity and its dependence on the dominant dust sources in late-type dwarf galaxies. Since the interstellar dust mixture is determined by the interplay of dust production by stellar sources and by dust growth and destruction in the ISM,  we use a model of the lifecycle of dust species of different origins, which has been used to study the dust content in the solar neighbourhood, Milky Way disk, and Large Magellanic Cloud (LMC) \citep{Zhukovska:2008p7215, Zhukovska:2008bw, Zhukovska:2013vg}. With this model modified for the evolution of dwarf galaxies, we re-evaluate the roles of individual dust sources in galaxies with various star formation histories. In particular, we establish when the dust growth in the ISM begins to dominate the dust production in model galaxies, and subsequent changes in the interstellar dust properties. 

The paper is organized as follows. Evolution models of dust and gas for dwarf galaxies are described in Sect.~\ref{sec:Mod}. Section~\ref{sec:Metal} presents the metallicity evolution in model galaxies and the resulting variations in the accretion timescales for dust growth in the ISM. In Sect.~\ref{sec:DustEv}, we present results of calculations of the DGR evolution,  verify our models by comparison with the relation between dust-to-gas ratio and oxygen abundance derived from observations, and discuss the dependence of our results on various galactic model parameters. Finally, conclusions are presented in Sect.~\ref{sec:Concl}.

\section{Model prescriptions}
\label{sec:Mod}

\subsection{Galactic evolution}
A model of the chemical evolution of the main dust-forming elements provides a good basis for the dust evolution model, since it naturally includes the abundance constraints for the interstellar dust mixture  \citep{ Dwek:1980p490, Dwek:1998p67}.  We adopt the model of the lifecycle of dust in the ISM introduced for the solar neighbourhood in  \cite{Zhukovska:2008bw}. Below we describe main modification needed for a dwarf galaxy model.

In contrast to spiral galaxies, dwarf galaxies have small sizes and no large metallicity gradients, so they can be modelled in one-zone approximation, with all quantities averaged for the whole galaxy. Therefore, equations of chemical evolution are solved for masses of elements in gas and dust instead of the surface densities used in \cite{Zhukovska:2008bw}. We assume that a model galaxy is formed by gas infall with the infall rate 
\begin{equation}
\dot{M}_{\rm inf} = \frac{M_{\rm G} {\rm e}^{-t/\tau_{\rm inf}}} { \tau_{\rm inf} (1 - {\rm e}^{-t_{\rm G}/\tau_{\rm inf}})}
\end{equation}
 on a short timescale of $\tau_{\rm inf}=$0.3~Gyr; $M_{\rm G}$ is the mass of gas accreted by infall for the whole time of galactic evolution $t_{\rm G}$, which is set to 13~Gyr in this paper. We assume that the infalling gas has a primordial composition (zero initial metallicity) following chemical and chemodynamical evolution models of dwarf galaxies \citep[e.g.,][]{Searle:1973fa, Recchi:2001bp, Recchi:2002p6651, Lanfranchi:2003p1954, Romano:2006db, Yin:2011he}. Alternatively, some chemical evolution models assume that the infalling gas is pre-enriched to  $Z\sim 10^{-4}-10^{-3}$ with type II SN-like enhanced $\rm [\alpha/Fe]$ ratio based on metallicities of damped Ly$\alpha$ systems \citep[e.g.,][]{Bekki:2012ge, Tsujimoto:2010uw}. We discuss how this assumption affects the model predictions in Sect.~\ref{sec:InfallAb}.
 
Star formation histories in late-type dwarf galaxies reconstructed from their colour-magnitude diagrams display a number of bursts of different durations separated by quiescent phases \citep[][ and references therein]{Tolstoy:2009cl, MartinManjon:2011cs}. \cite{Yin:2011he} find that the present-day observational constraints for chemical evolution of late-type dwarfs can be fitted with bursting models with no more than ten bursts and $\tau_{\rm SF}\sim 2$~Gyr. BCDs with continuous star formation should not be the majority of BCD population \citep{Yin:2011he}. Therefore, we adopt an episodic star formation in which long quiescent phases are separated by relatively short bursts of star formation occurring on a much shorter timescale $\tau_{\rm SF}$. The assumed star formation rate is linearly proportional to the gas mass
\begin{equation}
	\psi = \frac{M_g}{\tau_{\rm SF}}.
\end{equation}

\begin{table}[tb]
\caption{Main parameters of dwarf galaxy models}
\begin{tabular}{c c c c c c}
\hline\hline
\noalign{\smallskip}
Model &  $n_{\rm burst}$ & $t_{\rm burst} $& $dt_{\rm burst}$ & $\tau_{\rm SF}$ & $\tau_{\rm inf}$\\
 & & Gyr & Myr & Gyr & Gyr\\
\noalign{\smallskip}
\hline
Model 1 & 6 & 0, 1, 2, 5, 7, 11 & 50 & 2 &  0.3\\
Model 2 & 6 & 0, 1, 2, 5, 7, 11 & 500 & 2&  0.3\\
Model 3 & 6 & 0, 1, 2, 5, 7, 11 & 500 & 0.2 &  0.3\\
Model 4 & - & -& - & 10 & 0.3\\
Model 5 & 3 & 1, 5, 11 & 100 & 2 & 0.3\\
Model 6 & 5 & 0, 0.3, 0.6, 1, 7 & 50 & 0.2 &  0.3\\
Model 7 & 6 & 0, 1, 2, 5, 7, 11 & 50 & 2 & 1\\
\hline
\end{tabular}
\label{tab:Models}
\end{table}

We consider six starburst models with episodic star formation, which differ in the number of bursts  $n_{\rm burst}$, their duration $dt_{\rm burst}$, timescales of star formation $\tau_{\rm SF}$ and infall $\tau_{\rm inf}$ (Table~\ref{tab:Models}). It is assumed that all models are entering a starburst at the present time $t=13$~Gyr. The number of past starburst experienced at evolution times $t_{\rm burst}$ is listed in Table~\ref{tab:Models}. The burst duration is varied from 50 to 500~Myr. The star formation timescale during starbursts $\tau_{\rm SF}$ is set to 2~Gyr in all burst models, with exception of Model 3 and 6, for which it is reduced to 0.2~Gyr to study the effect of faster chemical enrichment. Model 3 does not represent a gas-rich starburst dwarf galaxy because it has a too short $\tau_{\rm SF}$  combined with a long starburst duration, resulting in the high metallicity and low gas fractions after initial starbursts. It is a model of evolved galaxy included for comparison. During the quiescence phases in starburst galaxies, $\tau_{\rm SF}$ is set to 200~Gyr. A non-zero star formation rate between starbursts is commensurate with the star formation histories of late-type dwarf galaxies derived from colour-magnitude diagrams \citep[][and references therein]{Tolstoy:2009cl}. The reference value for the infall timescale is 0.3~Gyr in all models except Model 7, for which the value of 1~Gyr is taken to illustrate the effect of slower galaxy formation. Model 4 with the continuous star formation on the timescale $\tau_{\rm SF} =$10~Gyr is only included for comparison.

\subsubsection{Galactic winds}
Galactic winds drive metals from the ISM and reduce its metallicity. Owing to their shallower gravitational potential wells, dwarf galaxies often experience galactic winds triggered by energetic stellar feedback \citep[see][for a recent review]{Recchi:2013vy}. In non-selective wind models, the entire ISM is affected by winds, and the gas loss is proportional to the star formation rate \citep[e.g.,][]{Bradamante:1998to, Lanfranchi:2003p1954, Calura:2008hi}.  Detailed dynamical models show that freshly produced metals from SNe are expelled by galactic winds more efficiently than the rest of the ISM \citep[differential winds, ][]{MacLow:1999p1955, Recchi:2004p6631}. Metal-enhanced galactic winds are required to explain recent chemical abundance constraints derived for dwarf galaxies \citep{Yin:2011he}.

In addition to dependence on galactic mass \citep[e.g.,][]{Bradamante:1998to}, the efficiency of galactic winds depends on many other galactic parameters, in particular, on luminosity,  efficiency of thermalisation of energy from type II and Ia SNe in the ISM, and geometry of the gas distribution \citep{Recchi:2001bp, Recchi:2013je}. To reduce the number of free parameters, we do not include galactic winds in the reference models and study their impact on dust content separately in Sect.~\ref{sec:GalWinds}. We follow \cite{Bekki:2012ge}, who adopt a selective wind model with the fixed fraction of expelled SN~II ejecta $f^{w}_{\rm  SN}$  treated as a free parameter. Hydrodynamic simulations of dwarf galaxy evolution indicate that metals from SN~Ia are more easily removed from the galaxy, because they are ejected into a hot rarefied bubble created by previous SNe \citep{Recchi:2001bp, Recchi:2002p6651, Recchi:2013je}. Therefore, in the present work, different values of $f^{w}_{\rm  SN}$ are chosen for SN~II and SN~Ia, $f^{w}_{\rm  SNII}=0.3$ and  $f^{w}_{\rm  SNIa}=0.6$, respectively.

\subsection{Dust model}
\label{sec:DustMod}
We adopt a multicomponent model following individual evolutions of grains of different origin (SNII, SNIa, AGB stars, and ISM-grown dust) \citep{Zhukovska:2008bw}. To account for the dust condensed in stellar winds of AGB stars, the mass- and metallicity-dependent dust yields are taken from calculations performed by \cite{Ferrarotti:2006p993} and \cite{Zhukovska:2008bw}. Dust input from AGB stars $Z \lesssim \Zs$ in these models is dominated by carbonaceous dust. 

For type II SNe, we adopt the condensation efficiencies (fractions of key chemical element ejected by all SNe in dust) from \cite{Zhukovska:2008p7215}, $\eta_{\rm sil} = 10^{-3}$,  $\eta_{\rm SiC} = 3 \times 10^{-4}$, and $\eta_{\rm car} = 0.15$ for silicate, SiC, and carbonaceous dust, respectively. These values should not be directly compared with the values derived from observations of dust condensation in SN ejecta, because the latter provides an upper value before part of dust is destroyed in reverse SN shocks. For condensation efficiencies adopted here SN dust is also dominated by carbon-rich grains, therefore the stardust mixture consists mainly of carbonaceous grains in our model. 
Since the efficiency of dust formation and survival in type II  SNe are still being debated, we investigate how much higher values of the condensation efficiencies, $\eta_{\rm sil} = \eta_{\rm SiC} = \eta_{\rm car} =0.5$ can affect the dust-to-gas ratio evolution in Sect.~\ref{sec:VarSN}.

\subsubsection{Dust destruction}
We follow \cite{Zhukovska:2008bw} to model the processing of grains (destruction and mantle accretion) in the ISM. Grains are destroyed in the ISM by SN blast waves on the timescale \citep{McKee:1989p1030}
\begin{equation}
	\tau_{\rm SNR} = \frac{M_{\rm ISM}}{m_{\rm cl} f_{\rm SN}R_{\rm SN}}\,,
	\label{eq:TauDestrGlob}
\end{equation}
where $m_{\rm cl}$ is the mass of gas that is completely cleared of dust by a single SN explosion, $f_{\rm SN}$ is the fraction of all SNe destroying dust, and $R_{\rm SN}$ the total rate of SNe. In contrast to spiral galaxies, SN Ia may be important for the dust destruction during quiescent phases in a dwarf galaxy with episodic star formation. The $f_{\rm SN}$ is introduced to account for correlated SNe that explode in a bubble created by previous SNe and therefore are less efficient in dust destruction. We assume that they destroy 10\% of the dust compared to a single SN \citep{McKee:1989p1030}. We adopt the value of $f_{\rm SN}=0.35$ derived by \cite{Zhukovska:2013vg} using the observed fraction of field OB stars in the LMC, a dwarf galaxy in the phase of active star formation. For $m_{\rm cl}$, we adopt the values of 1300 and 1600\Ms\, which were also estimated for the LMC in \cite{Zhukovska:2013vg}.

\subsubsection{Dust growth in the ISM}
We assume that dust growth occurs by selective accretion of gas species in dense clouds on existing grains in the sense that carbon atoms stick to the carbonaceous grains and Si, Fe, Mg, and O stick to the silicate grain surfaces \citep{Draine:2009p6616}. A model for the dust growth adopted here is described in \cite{Zhukovska:2008bw}, and we direct the reader to the original publication for details. 
Dense clouds in our model are characterised by the temperature $T_{\rm cl}$,  the particle number density $n$, the mass fraction relatively to the total gas mass $X_{\rm cl}$, and lifetime $\tau_{\rm exch}$. The $\tau_{\rm exch}$ is also a timescale of exchange between dense and diffuse components of the ISM.
 
Grain growth is determined by collisions of some key element (growth species) with grain surfaces. The equation for the condensation degree of a dust-forming species in a dense cloud is
\begin{equation}
	\ddt{ f} = {f(1-f)\over \tau_{\rm acc}}\,,
\end{equation}
where $\tau_{\rm acc} $ is the the accretion timescale, and it is the key quantity for dust growth. For the dust species of type $j$, it is 
\begin{equation}
           \tau^{-1}_{j,\rm acc} = {3 \alpha_j A_{j} m_{\rm amu}\over \rho_{j,\rm c} \nu_{j,\rm c}} \cdot {1\over \langle a_j \rangle} \cdot \upsilon _{j,\rm th} n_{\rm H}   \cdot \epsilon_j \, ,
\label{eq:TauGrowthDu}
\end{equation}
where $\alpha_j$ is the sticking efficiency to the grain surface, 
$A_{j}$ the atomic weight of one formula unit of the dust species,
$\rho_{j,\rm c} $ is the density and $\nu_{j,\rm c} $  the number of atoms of the key element contained in the formula unit of the condensed phase, 
$\langle a_j\rangle$ the average grain radius,
$\upsilon _{j,\rm th}$ the thermal speed of the growth species, 
$n_{\rm H}$ the number density of H nuclei in clouds,
and $\epsilon_j$ the element abundance of the key species.

Because only a small fraction of the ISM resides in dense gas in dwarf galaxies \citep[e.g.,][]{Leroy:2005ke}, many timescales are required to cycle all ISM through clouds on an "effective" exchange time
\begin{equation}
\tau_{\rm exch,eff} =\tau_{\rm exch}{1-X_{\rm cloud}\over X_{\rm cloud}}\,.
\label{Eq:EffTimesc}
\end{equation}
The dust production rate is
\begin{equation}
G_{j,\rm d}={1\over\tau_{\rm exch,eff}}\Bigl[\,f_{j,\rm ret}M_{j,\rm d,max}-
M_{j,\rm d}\,\Bigr]\,,
\label{eq:DuProdInCloud}
\end{equation}
where $f_{j,\rm ret}$ is the condensation degree for growth species of kind $j$ after cloud dispersal, $M_{j,\rm d,max}$ is the maximum possible dust mass assuming complete condensation, and $M_{j,\rm d}$ the current average dust mass. For $f_{j,\rm ret}$, we adopt an approximated formula given by Eq.~(45) in \cite{Zhukovska:2008bw}
\begin{equation}
     f_{j,\rm ret}(t) =\left[  \left( f_{j,0}(t) ( 1+ \tau_{\rm exch}/\tau_{\rm acc} ) \right)^{-2}
      +1  \right]^{-{1/2}}\,,
\label{eq:fret}
\end{equation}
where $f_{j,0}$ is the initial degree of condensation for species $j$. 

The growth timescale $\tau_{j,\rm acc}$ given by Eq.~(\ref{eq:TauGrowthDu}) is connected to the chemical enrichment of the cold phase through the element abundance of the key element $\epsilon_j$. We consider growth in the ISM of three dust species (silicate, carbonaceous, and iron grains) and adopt the same values for the condensed phase parameters as described in \cite{Zhukovska:2008bw}.

\begin{figure}[tb]
	        \centering
	     	\includegraphics[width=0.5\textwidth]{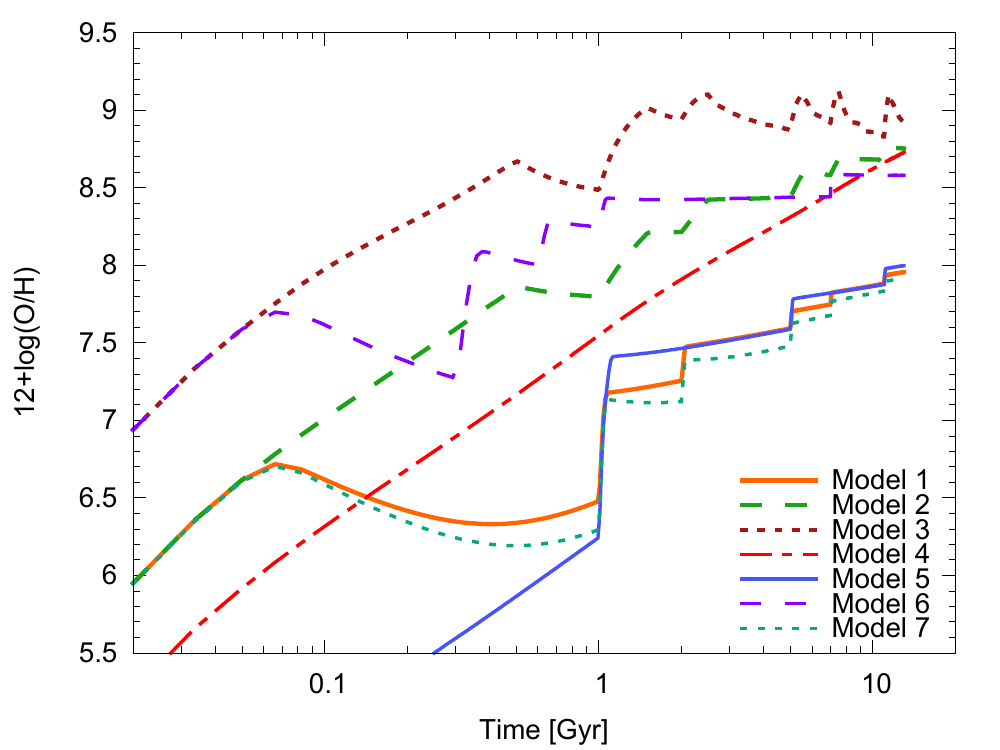} 
			\caption{Time evolution of oxygen abundance for different dwarf galaxy models (Table~\ref{tab:Models}). Depending on the star formation history, the same value of O abundance can be reached at very different instants.}
			\label{fig:O-t}
\end{figure}

For the grain size distribution, we assume a power-law distribution $a^{-q}$ with the index $q=3.5$ with the minimum and maximum grain sizes $a_{\min}=0.005\mum$ and $a_{\max}=0.25\mum$, respectively \citep[][hereafter MRN]{Mathis:1977p750}, which yields $\langle a \rangle=0.035\mum$. \cite{Nozawa:2013dr} have recently confirmed that a simple MRN graphite-silicate grain mixture with $a_{\max}=0.24\pm0.05\mum$ reproduces the extinction curves of both the Milky Way and Small Magellanic Cloud, which led the authors to suggest that the power law $a^{-3.5}$ is a generic power law expected from the collisional equilibrium. \cite{Kuo:2012ej} note that variations in the grain size distribution can be important for the dust growth in the ISM.  Here we do not consider the effect of these variations  and restrict our study to the impact of various galactic parameters and the chemical evolution on the dust content.

Dwarf galaxies exhibit the multiphase ISM structure dominated by neutral gas. There are a few detections of molecular gas in metal-poor dwarfs, because its main tracer, the CO molecule, is very difficult to detect at low metallicities \citep[e.g.,][]{Leroy:2005ke, Leroy:2009gj, Schruba:2012bb}. For $X_{\rm cl}$ in reference models, a value of 0.10 is adopted based on the ratio of median values of \ion{H}{I} and H$_2$ masses in the galaxy sample in \cite{Leroy:2005ke}. However, the molecular gas fraction should be higher during a starburst and decreases afterwards due to stellar feedback \citep[e.g.,][]{Melioli:2004di}. 
 \cite{McQuinn:2012io} indirectly confirm the presence of significant amounts of molecular gas in nearby starburst dwarf galaxies at the onset of the starbursts required to explain high stellar surface densities and relatively low surface densities of atomic gas. Therefore, we inspect how variations in $X_{\rm cl}$ affect the dust growth in Sect.\ref{sec:CloudVar}, assuming that $X_{\rm cl}$ varies from 0.01 during quiescent phases up to 0.5 during starbursts. 
For $\tau_{\rm exch}$, a value of 10~Myr is adopted. Our reference value for physical conditions in clouds are 50~K for $T_{\rm cl}$ and $10^{3}\cmc$ for $n_{\rm H}$.

%
\begin{figure}[tb]
			\includegraphics[width=0.5\textwidth]{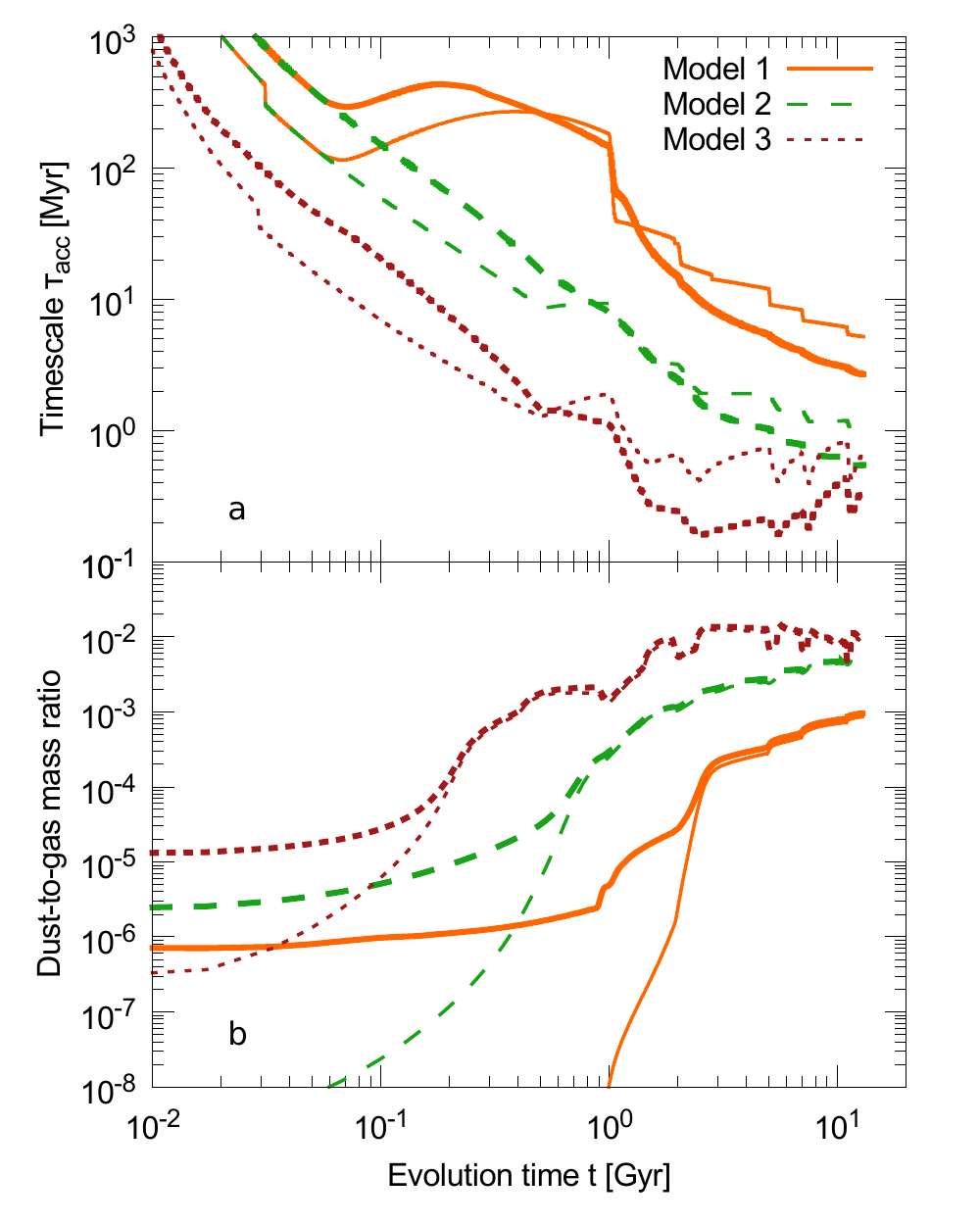}
			\caption{(a) Time evolution of accretion times scales for silicate and  carbonaceous grains (thin and thick lines) for Models 1, 2, and 3 (solid, long-dashed and short-dashed lines, respectively). (b) Dust-to-gas mass ratio (DGR) evolution with time from the onset of galaxy formation for Model 1--3 shown with the same line types as in the top panel. Thick and thin lines show the total DGR and the DGR calculated with only the ISM-grown  dust.}
			\label{fig:DGR-tau-t}
\end{figure}

\begin{figure}[tb]
			\includegraphics[width=0.5\textwidth]{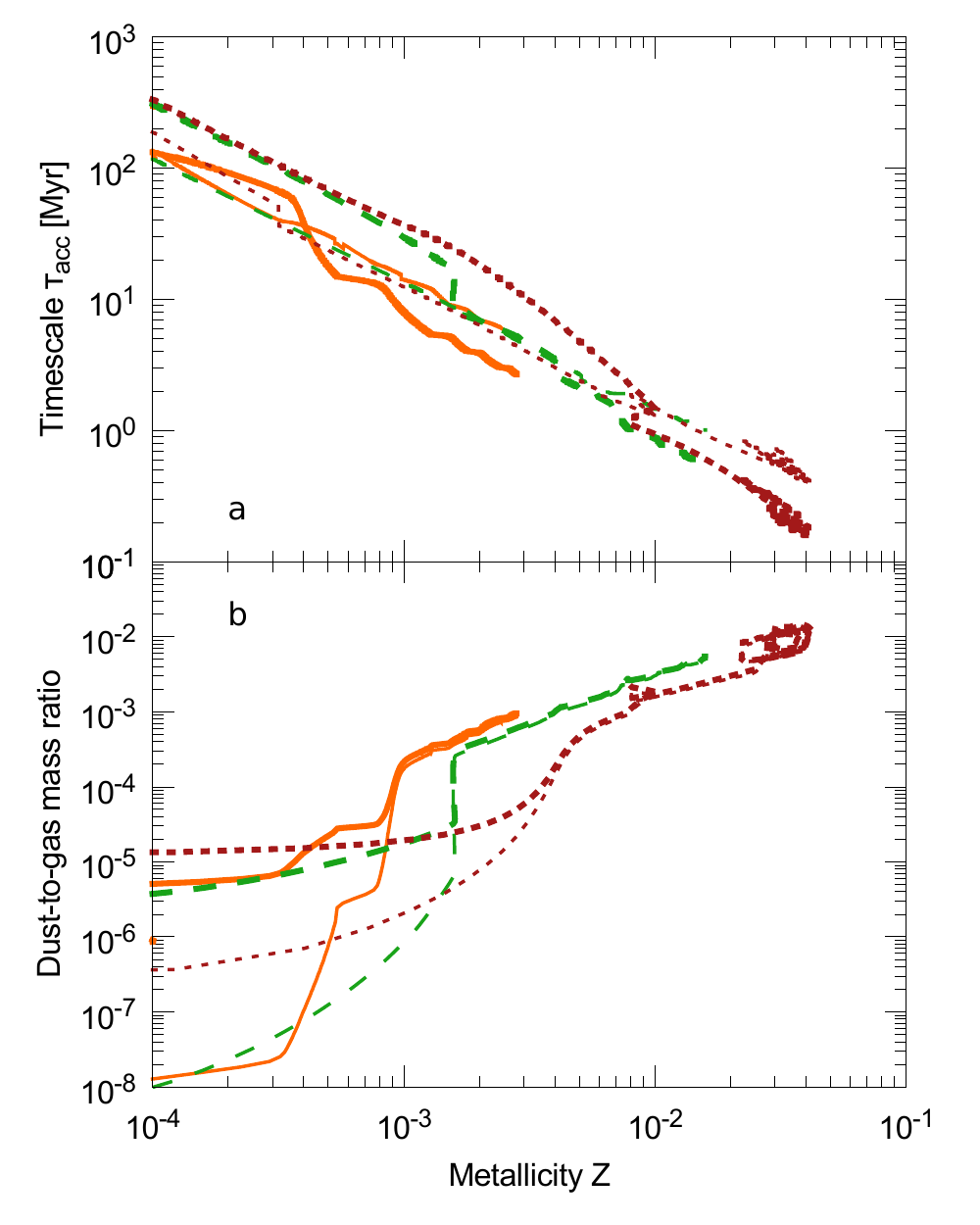}
			\caption{The same as in Fig.~\ref{fig:DGR-tau-t} as a function of metallicity.}
			\label{fig:DGR-tau-Z}
\end{figure}

\section{Metallicity and accretion timescales}
\label{sec:Metal}
Oxygen is the most abundant element in galaxies after H and He, and it is the main constituent of the silicate dust. The abundance of oxygen is easier to determine than for other heavy elements, so it is commonly used as a metallicity indicator. 

The results of calculations of the oxygen abundance evolution for model dwarf galaxies (Table~\ref{tab:Models} ) are shown in Fig.~\ref{fig:O-t}. The O abundance undergoes a complex evolution in galaxies with bursting star formation in contrast to Model 4 with continuous star formation. Since oxygen is mainly produced by massive stars, its abundance increases stepwise during each star formation burst. In Models 1, 2, and 3,  the O abundance in the ISM decreases between the first and the second bursts of star formation owing to the infall of primordial gas. 
In Model 7, which is identical to Model 1 with the exception of the longer infall timescale $\tau_{\rm inf}$, a dip in the O abundance is only 0.2 dex higher than in Model 1. This difference is reduced in later enrichment episodes, demonstrating a weak dependence of the ISM metallicity on our choice of $\tau_{\rm inf}$ value.
At later evolution times, the infall rate is negligible, and oxygen evolution is determined by the duration and strength of bursts. In Model 1, it increases between bursts owing to the quiescent star formation. Interestingly, in Model 3, \OH\ decreases during the quiescent phases because of the mass return from metal-poor low-mass stars formed during early galactic evolution at lower metallicities. Metallicity in their envelopes is not noticeably modified by dredge-ups during post-main sequence evolution, therefore their mass loss reduces the total ISM metallicity in the presented highly evolved model. It is not the case for a typical late-type dwarf, though.

Galactic chemical evolution determines the variations in the accretion (growth) timescales $\tau_{\rm acc}$ controlling  the dust production rates in the ISM. Time variations of $\tau_{\rm acc}$ for silicate and carbon dust for Models 1--3 are shown in Figs.~\ref{fig:DGR-tau-t}a. Growth timescales for silicate and carbonaceous grains demonstrate different behaviours: $\tau^{\rm sil}_{\rm acc}$ evolution reflects stepwise changes in metallicity with each starburst (Fig.~\ref{fig:O-t}), while $\tau^{\rm car}_{\rm acc}$ decreases smoothly on a longer timescale because of the delayed carbon production by low-mass stars. Initially, $\tau^{\rm car}_{\rm acc} \approx 3 \tau^{\rm sil}_{\rm acc}$, but after 1~Gyr in Model 1 and 2 and  300~Myr in Model 3, $\tau^{\rm car}_{\rm acc}$ becomes shorter than $\tau^{\rm sil}_{\rm acc}$. Efficient growth in the dense clouds is expected, when $\tau_{\rm acc}$ is smaller than $\tau_{\rm exch, eff}$ given by Eq.~(\ref{Eq:EffTimesc}), which equals to 90~Myr for our reference models.

Figure~\ref{fig:DGR-tau-Z}a clearly demonstrates that, in contrast to $\tau^{\rm sil}_{\rm acc}$, the growth timescales for carbon grains vary significantly for the same values of metallicities depending on the SFH because of the delayed enrichment of the ISM with carbon by long-lived low-mass stars.

\section{Results/discussion}
\label{sec:DustEv}
In the following we present results of  the calculations of dust evolution in the model dwarf galaxies with various star formation histories (Table~\ref{tab:Models}). The dominant dust sources and corresponding values of the dust-to-gas ratio during galactic evolution are discussed in detail for Models 1 to 3, which encompass a wide range of star formation rates.  Then, we compare the relation between the dust-to-gas ratio and oxygen abundance for all models with that derived from observations of a dwarf galaxy sample.

\subsection{Dust-to-gas ratio evolution and dust sources}

Evolution of the dust-to-gas mass ratio (DGR) with time and metallicity  for Models 1 to 3 is shown in Figs.~\ref{fig:DGR-tau-t}b and \ref{fig:DGR-tau-Z}b, respectively. The figures display both the total DGR and the DGR due to the ISM-grown dust component.

A common feature of the models in Fig.~\ref{fig:DGR-tau-t}b is a plateau at DGR$\lesssim 10^{-5}$ at early epochs (low metallicities) followed by a steep rise due to the increasing importance of dust growth in the ISM. When dust-forming elements (Si, Mg, Fe) in the ISM are  almost completely condensed in dust, the growth process reaches saturation. For continuous star formation, \cite{Asano:2013kl} points that in the saturation regime, various evolution models converge in DGR vs. Z plot to a simple linear dependence. The value of the dust-to-metal ratio is then determined by the fraction of dust-forming elements available in the ISM, which does not strongly depend on the chemical evolution and is close to the 0.4 found in spiral galaxies \citep{Dwek:1998p67, Draine:2007de, James:2002gj}. Models of dust growth adopted here combined with bursting star formation scenarios also converge to the linear dependence of the DGR on metallicity (Fig.~\ref{fig:DGR-tau-Z}b), but the moment when it happens depends not only on the amount of metals as in models with continuous star formation \citep{Asano:2013kl}, but also on post-burst galactic evolution.

\begin{figure}[tb]
			\includegraphics[width=0.5\textwidth]{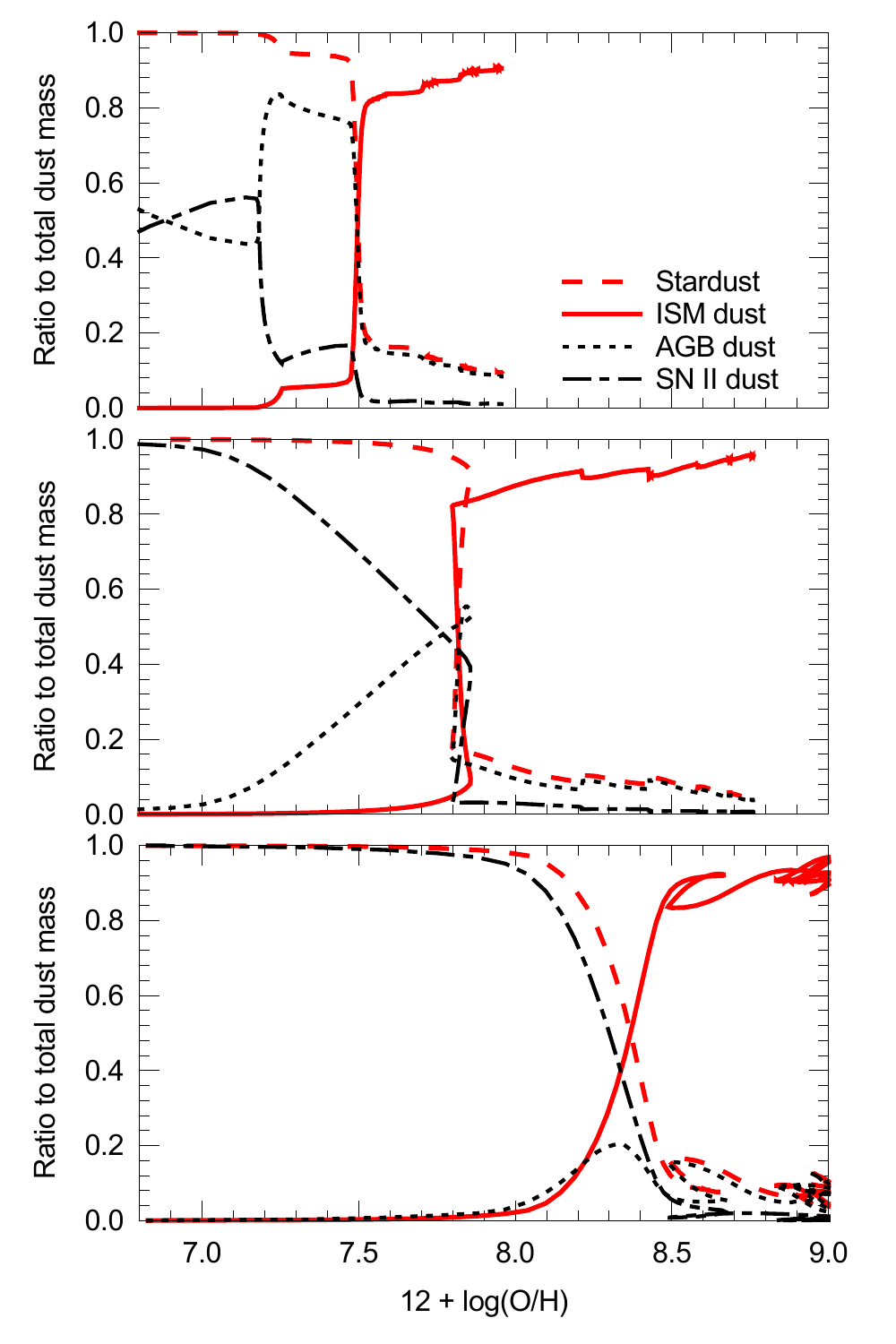}
			\caption{Mass fractions of dust grown in the ISM (solid lines),  dust from AGB stars (short dashed), type II SN (dot dashed), and total stardust (AGB+SNII, long-dashed) relatively to the total dust mass. Top, middle, and bottom panels show results for Models 1, 2, and 3, correspondingly (Table~\ref{tab:Models}).}
			\label{fig:Ratio-O}
\end{figure}

The DGR evolution in model dwarf galaxies demonstrates stepwise changes with metallicity, most clearly seen in Model 1 representing a typical gas-rich dwarf galaxy (Fig.~\ref{fig:DGR-tau-Z}b). Horizontal steps correspond to a fast enrichment by massive stars during starbursts and vertical steps to an increase in the DGR between starbursts cause by either dust input from AGB stars or dust growth in the ISM. When the ISM dust growth saturates, the DGR increases more smoothly with $Z$ approaching a linear relation, because most existing metals are locked up in dust and accretion of freshly produced metals by starburst does not significantly change the DGR value. Deviations of the DGR from this linear dependence in saturation regime, the most pronounced in Model 3 (Fig.~\ref{fig:DGR-tau-Z}b), are caused by efficient dust destruction during bursts. Loops in DGR vs. $Z$ plot present in Model 3 near $Z=0.001$ and $Z\gtrsim 0.02$ stem from alternated operations of dust growth in the ISM during quiescent phases and efficient dust destruction during starbursts. Such intense dust destruction is intrinsic to evolved galaxy models with low gas fraction, and is not expected for metal-poor gas-rich galaxies.

\subsubsection*{Critical metallicities for dust growth in the ISM}
Figures~\ref{fig:DGR-tau-t}b and \ref{fig:DGR-tau-Z}b reveal that the transition from stardust- to ISM growth-dominated dust production occurs at very different times and metallicities: at instants $t=200$~Myr in Model 3, $t=600$~Myr in Model 2, and only  $t=2$~Gyr in Model 1 corresponding to  metallicities of $3\times 10^{-3}$, $1.6\times 10^{-3}$, and $9\times 10^{-4}$. Such differences can be understood from analysis of the accretion timescale variations with time and metallicity (Figs.~\ref{fig:DGR-tau-t}a and \ref{fig:DGR-tau-Z}a).  For silicates, the value of $\tau^{\rm sil}_{\rm acc}$ is very similar for three models for a given metallicity, while the time needed for the ISM to get enriched to this metallicity is the longest for Model 1 and the shortest for Model 3. Thus, metals can accrete on grains during a significantly longer period until metallicity is changed to a higher value in Model 1 compared to Model 3. As a result, the value of the dust-to-gas ratio due to the dust growth in the ISM for a given $Z$ value is higher in galaxies with slower enrichment than in galaxies with more rapid enrichment.

\begin{figure}[t]
			\includegraphics[width=0.5\textwidth, page=1]{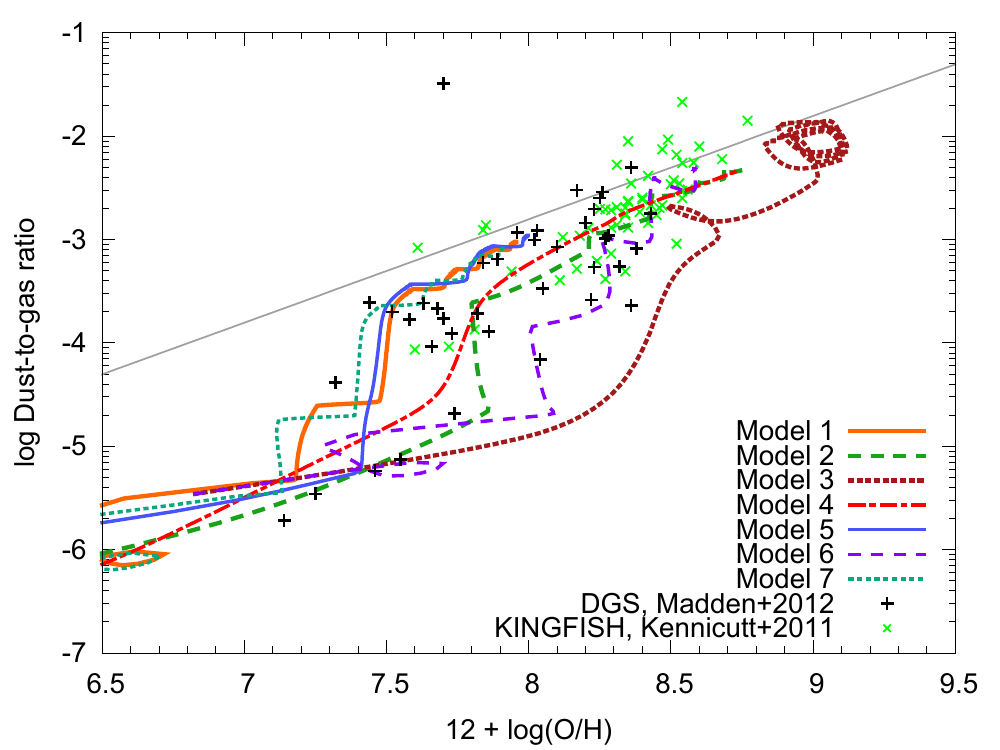} 
			\caption{Variations in the dust-to-gas mass ratio with oxygen abundance for model dwarf galaxies (Table~\ref{tab:Models}). Different symbols show the values derived from observations \citep{RemyRuyer:2013uu}. The grey line shows a linear relation between the DGR and O abundance derived for the solar neighbourhood.}
			\label{fig:DGR-O}
\end{figure}

\subsection{Roles of stellar sources}\label{sec:RolesStars}
Variations in the dust mass fraction from AGB stars, type II SNe, and ISM relative to the total dust mass in Models 1--3 are shown as a function of  the oxygen abundance in Fig.~\ref{fig:Ratio-O}.
AGB stars and SNe II play various roles in dust production for the same values of metallicities in considered models.  At metallicities below the critical metallicity in Model 1, dust input is mostly dominated by AGB stars, while type II SNe are only important during the first starburst and briefly during the second (Fig.~\ref{fig:Ratio-O}, top). In Models 2 and 3, the dust growth in the ISM overtakes the dust production before the moment, when the dust input from AGB stars exceeds dust input from SN~II, and AGB stars never become the dominant dust sources. 

Thus, AGB stars formed during starbursts become important dust factories if the accretion timescale remains longer than the effective timescale of the matter cycle between dense and diffuse gas for sufficiently long periods comparable to the lifetimes of low- and intermediate-mass stars. 
This can be the case in metal-poor late-type galaxies, which host underlying old stellar populations with ages of $\gtrsim 1$~Gyr \citep{Tolstoy:2009cl, Kunth:2000p6606}.

\subsection{Comparison with observations}

\subsubsection{Observational data}
The relation between the dust-to-gas ratio and metallicity derived from observations provides a valuable test for models of dust evolution. The dust-to-gas ratio in dwarf galaxies has been studied observationally in many works over the past decades \citep[e.g.,][]{Lisenfeld:1998p491, Walter:2007cj, Grossi:2010p6588, James:2002gj, Hunt:2005ks, Fisher:2013tp, Galametz:2011kxa, Engelbracht:2008p4082}.  Dust masses are usually estimated from dust emission at far-IR and/or submillimetre wavelengths. The total gas mass is calculated by adding atomic and molecular hydrogen gas masses and correcting afterwards for contributions of He and heavier elements. Mass of \HI gas can be directly calculated from the intensity of the 21~cm line. The \H2 mass is estimated indirectly from CO emission, which is coupled with intrinsic difficulties in low-metallicity systems ($\OH \lesssim 8.0$) owing to extremely faint CO emission and uncertainties in the CO-to-\H2 mass conversion \citep[e.g.,][]{Schruba:2012bb}. Detection of dust emission is also limited by sensitivity at low metallicity \citep[][and references therein]{Madden:2013iu}.

Recently, the high sensitivity of \textit{the Herschel Space Observatory} has significantly improved the accuracy of DGR measurements in metal-poor dwarf galaxies and allowed to access undetected dust emission in systems within the Dwarf Galaxy Survey (DGS) \citep{Madden:2013iu, ARemyRuyer:2013hr}. In the present work, we compare our models with the most recent determinations of the DGR based on the  DGS sample derived by \cite{RemyRuyer:2013uu}. They collected both IR and submm data from various instruments and fitted them with the dust SED model from \cite{Galliano:2011gy} to determine the dust masses. For gas mass estimates, they compiled data from the literature with a correction for more extended \HI distribution compared to the aperture that probed dust in dwarf galaxies and for undetected molecular gas.

\begin{figure}[t]
			\includegraphics[width=0.5\textwidth, page=2]{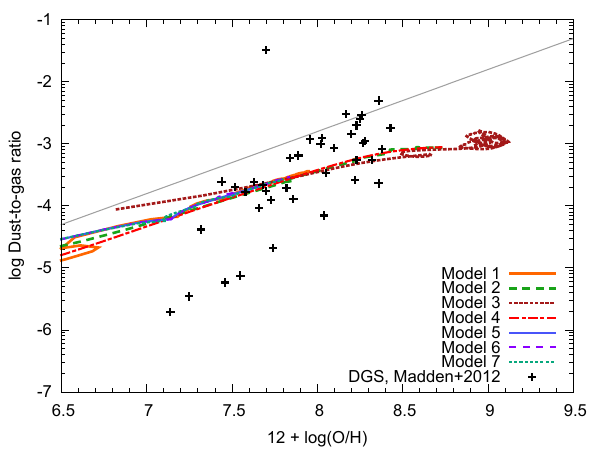} 
			\caption{The same as in Fig.~\ref{fig:DGR-O}, but calculated with the enhanced dust input from type II SNe without dust growth in the ISM.}
			\label{fig:DGR-O_SN}
\end{figure}

\subsubsection{ISM growth vs. enhanced dust input from type II SNe}\label{sec:VarSN}
As discussed above, our models of dust evolution in dwarf galaxies show that dust growth in the ISM can play an important role in these objects. 
In these calculations, we assume relatively low efficiencies of dust condensation in type II SNe. Recent far-IR and submillimetre observations revealed, for the first time, higher masses of newly-formed dust in SN ejecta \citep{Matsuura:2011ij, Gomez:2012fm, Lakicevic:2011fl}. While it is still unclear what fraction of freshly condensed dust in SN ejecta will survive the passage of reverse shock, the observed relation between DGR and metallicity can provide a test for the dominant mechanism for dust formation.

To check whether the observed variations in the DGR with metallicity can be explained without dust growth in the ISM simply by enhanced dust input from SNe~II, we perform model calculations with condensation degrees of 0.5, which is much higher than the values used in the reference models described in Sect.~\ref{sec:DustMod}. Results of the DGR evolution with oxygen abundance for Models 1 to 7 for the standard dust model and enhanced SN~II dust production without the ISM growth are  shown in Figs.~\ref{fig:DGR-O} and \ref{fig:DGR-O_SN}, respectively. In both figures the model predictions are compared with the DGR values for galaxies from DGS derived by \cite{RemyRuyer:2013uu}. Figure~\ref{fig:DGR-O} also shows the DGR values calculated by \cite{RemyRuyer:2013uu} for the KINGFISH sample, which includes metal-rich galaxies (spiral, early-type, and a few irregular galaxies), for comparison.

The standard models with low condensation efficiencies in type II SNe and dust growth in the ISM reproduce the observed relation between DGR and O abundance well, from the lowest values up to  $\OH \sim 8.4$ (Fig.~\ref{fig:DGR-O}).  
The locations of the most metal- and dust-poor galaxies on the DGR--O plot coincide with early evolution epochs of model galaxies, when dust input is dominated by stellar sources. The most metal-poor galaxy, I~Zw~18, lies slightly below the model predictions. One of possible reasons for this is the non-zero metallicity of the infalling material, resulting in somewhat lower DGR values at low metallicities than in the standard models, which will be discussed in Sect.~\ref{sec:InfallAb}. 
A large dispersion in the observed values of the DGR for $7.2 < \OH <8.5$ can be explained by the increasing importance of dust growth by accretion in the ISM, which occurs at various critical times and metallicities depending on the SFH. For example, the most metal-poor galaxies ($\OH<7.5$) with higher values of the DGR are described by Models 1, 5, and 7 characterised by slow chemical enrichment. They can be much older ($\gtrsim 1$~Gyr) than galaxies with the DGR values of a few $10^{-6}$. Dust-deficient galaxies with $\OH \gtrsim 8$ are best described by Model~6, with a short $\tau_{\rm SF}$ and shorter time intervals between bursts (300~Myr) compared to other models. Model~3 with the same value of $\tau_{\rm SF}$ and longer starburst/post-starburst durations results in an extremely fast chemical enrichment, which occurs faster than the dust growth in the ISM. Therefore, the DGR values predicted by this model are lower than values determined from observations. This model is only included for illustration purpose.
At higher metallicities,  our models reproduce the general trend, but cannot explain the values above the line corresponding to complete condensation, in particular for KINGFISH galaxies. These high (supersolar) values of the dust-to-metals ratio are difficult to reconcile with the present dust model, and it is beyond the scope of this paper. 

\begin{figure}[t]
			\includegraphics[width=0.5\textwidth]{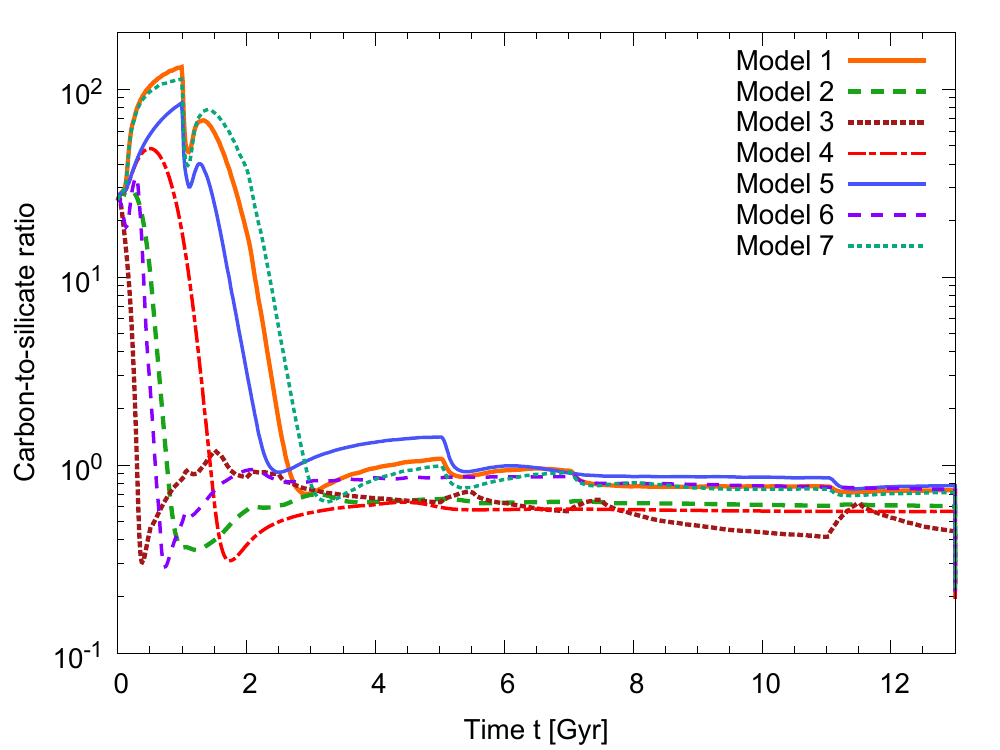} 
			\caption{Time variations of carbon-to-silicate dust mass ratio in model galaxies with different star formation histories.}
			\label{fig:carbsil-t}
\end{figure}

In models with enhanced dust input from type II SNe and no dust growth in the ISM, the predicted DGR at low metallicities is an order of magnitude higher than the values determined from observations (Fig.~\ref{fig:DGR-O_SN}). Moreover, the resulting slope is very similar for different models and shallower than the trend of observational data. In contrast to AGB stars, dust production in type II SN only weakly depends on metallicity. Thus, to  reproduce the observed variations of the DGR--O relation without dust growth in the ISM, very high condensation efficiencies in SNe are required by the high-metallicity part of the diagram, and a very efficient mechanism of dust removal/destruction is required to describe the low-metallicity part. It is unlikely that more than 99\% of  grains from SN II are destroyed in starbursts at low metallicities, but they survive starbursts almost intact at higher Z. Thus, the observed  DGR--O relation is best reproduced by models with a low net production by SNe type II, and an additional increase in the dust mass in the ISM.

The main difference between dust production dominated by type II SNe and ISM growth is that the latter increases dust content \textit{between the bursts}, while type II SNe produce dust (and raise the DGR) only during the star-formation bursts. \cite{Grossi:2010p6588} have investigated the DGR in dwarf galaxies, which are forming stars less actively than typical BCDs, and found that it is about ten times higher than the DGR values in galaxies with active star formation. For example, the value of the DGR of $3-6\times 10^{-4}$ in the dwarf irregular galaxy NGC 1569 is indeed much lower than in the more quiescent galaxies studied by \cite{Grossi:2010p6588}.

\subsection{Carbon-to-silicate dust mass ratio}
With our current assumptions for stardust input, the dust mixture from both type SNe and AGB stars is dominated by carbon-rich dust. Therefore, in early epochs, the carbon-to-silicate dust ratio is characterised by much higher values compared to the present-day value of 0.52 predicted for the solar neighbourhood \citep{Dwek:2005p1018}. Since the ISM chemistry is O-rich,  transition to the dust production by the ISM growth is marked by a strong variation in the carbon-to-silicate dust ratio in the ISM (Fig.~\ref{fig:carbsil-t}).  The ratio decreases from about 100 to $\sim 0.5-1$, when the ISM growth starts to dominate the dust production. This drastic change in the carbon-to-silicate ratio predicted by our model can be used as an indicator of the transition to the dust production by growth in the ISM. However, since the efficiency of dust condensation in SN type II is still debated, and SN may produce larger amounts of silicate grains than assumed in the present work, the initial ratio of carbon-to-silicate dust can be lower.

\begin{figure}[t]
			\includegraphics[width=0.5\textwidth, page=1]{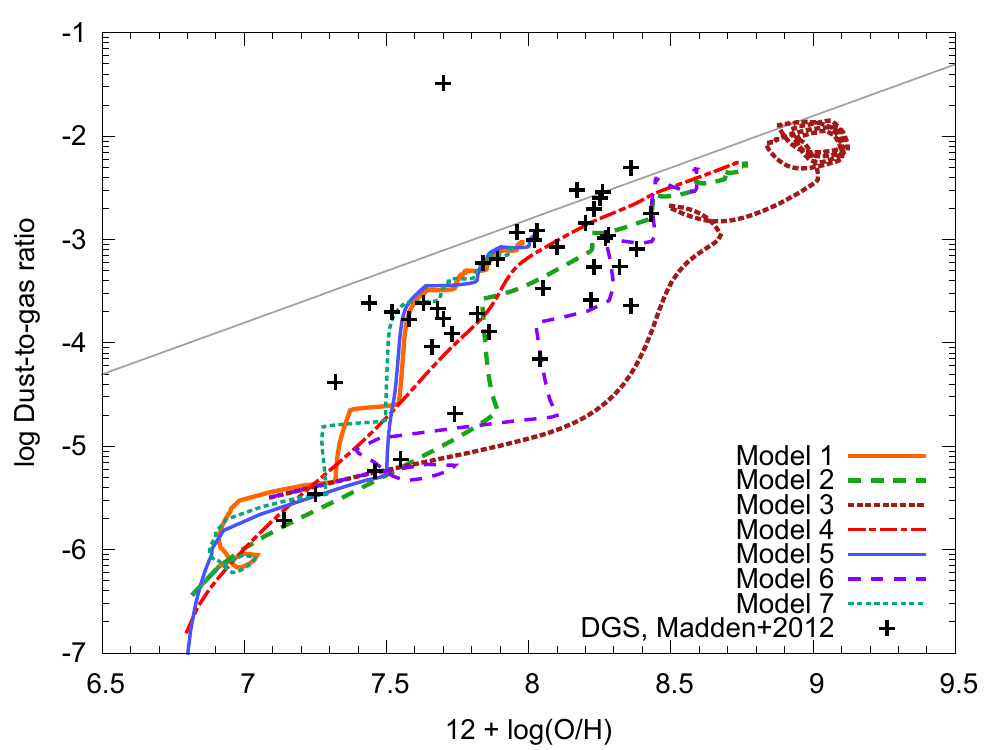} 
			\caption{The same as in Fig.~\ref{fig:DGR-O}, but calculated with galaxy formation by pre-enriched infall with the metallicity $Z=10^{-4}$ and SN~II like element ratios.}
			\label{fig:DGR-O_Zinf}
\end{figure}

\subsection{Dependence on galaxy parameters}\label{sec:DependParam}

\subsubsection{Infall abundance}\label{sec:InfallAb}
We performed model calculations for the same set of models (Table~\ref{tab:Models}), but instead of primordial infall, we assumed that model galaxies are formed by infall of pre-enriched gas with $Z=10^{-4}$ (1/140\Zs) and element ratios as in SN~II ejecta \citep{Bekki:2012ge, Tsujimoto:2010uw}. Infalling gas was assumed to be dust-free. The resulting variations in the DGR with oxygen abundance are presented in Fig.~\ref{fig:DGR-O_Zinf}, together with values from observations obtained by \cite{RemyRuyer:2013uu}.

As expected, the difference between the pre-enriched and reference models shown in  Fig.~\ref{fig:DGR-O} appears only at low O abundances comparable to the value of $\OH=6.8$  in the infalling gas. Pre-enriched dust-free infall results in a steep drop in the DGR for  $\OH\lesssim 7.2$, in contrast to a much flatter evolution in the models with primordial initial abundances. At later times, the impact of pre-enriched infall manifests in a shift in the DGR in Models 1 and 5 by about 0.1~dex towards higher O abundances relative to the reference models. Since dust-forming elements are initially available in the ISM, these model galaxies are enriched faster, resulting in higher critical metallicities. Thus, the DGR values are lower than in the reference models, for the same metallicities, until dust growth by accretion reaches saturation in both sets of models (Sect.~\ref{sec:DustEv}). Metallicity in Models 6 and 3 with fast chemical enrichment rapidly exceeds the initial metallicity of the infall, therefore the DGR evolution in these models is hardly affected by the infall abundances.

It seems that a steep decrease and small scatter in the DGR at low metallicities predicted by models with pre-enriched infall reproduces the observational data better, in particular, the DGR values in two most metal-poor galaxies, I Zw 18, and SBS0335-052, derived by \cite{RemyRuyer:2013uu} (Fig.~\ref{fig:DGR-O_Zinf}). However, it is hard to make the final conclusion that the observational data favour non-zero infall metallicity because of a small number of objects observed at these extremely low metallicities.

\begin{figure}[t]
			\includegraphics[width=0.5\textwidth, page=1]{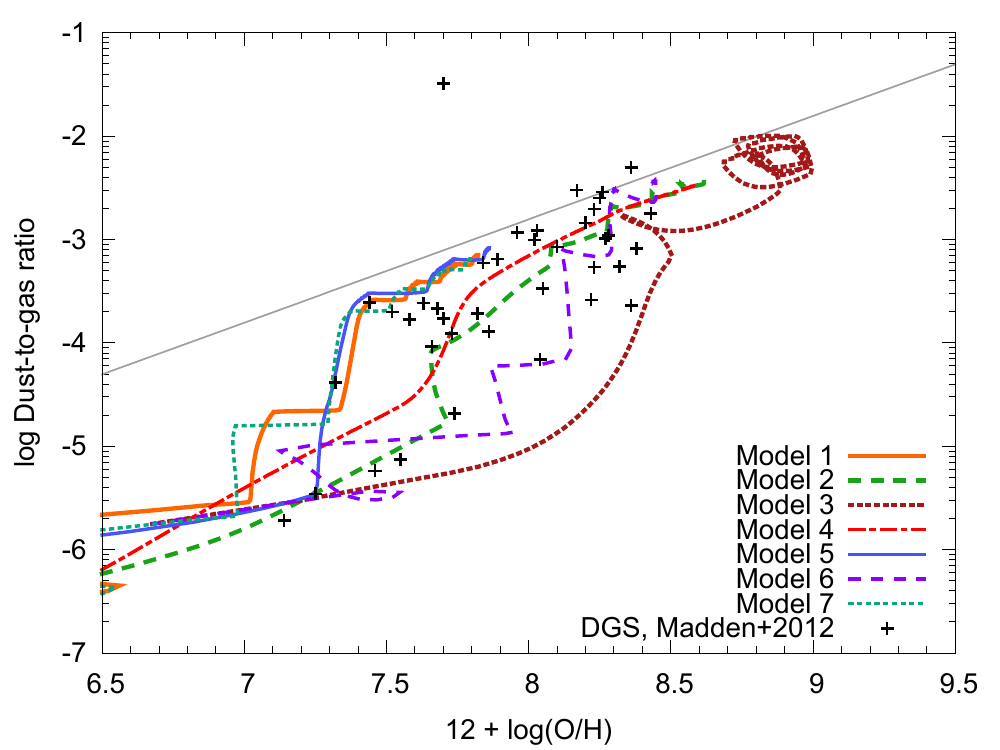} 
			\caption{The same as in Fig.~\ref{fig:DGR-O}, but calculated with selective galactic winds expelling 30\% and 60\% of SN~II and SN~Ia ejecta, respectively.}
			\label{fig:DGR-O_wind}
\end{figure}

\subsubsection{Galactic wind}\label{sec:GalWinds}
We performed calculations of dust evolution for model galaxies listed in Table~\ref{tab:Models} with selective galactic winds, which preferentially remove ejecta from type II and Ia SNe with the corresponding fractions $f^{w}_{\rm SNII}=0.3$ and $f^{w}_{\rm SNII}=0.6$. Although some fraction of the ambient ISM is entangled by galactic outflows and removed from the galaxy, dynamical models show that it is relatively low \citep[e.g.,][]{MacLow:1999p1955}.  We can thus assume that only SN ejecta are expelled from the galaxy. In this case, the effect of winds on total gas mass and SFH in galaxies is negligible. It means that  the SN rate and dust destruction timescales have the same values as in the reference models. Dust production by type II SNe is hardly affected by selective winds, since nucleosynthesis yields for SNe~II weakly depend on metallicity. The dust mixture from AGB stars is more sensitive to metallicity \citep[e.g.,][]{Ferrarotti:2006p993}. A decrease in the ISM metallicity leads to an increase in carbon-to-silicate dust mass ratio in stellar outflows from AGB stellar populations, although these changes will probably not be evident at low metallicities, at which AGB stars are the dominant dust source (Sect.~\ref{sec:RolesStars}). Also dust growth in the ISM strongly depends on metallicity as explained in Sect.~\ref{sec:DustEv}. In fact, the effect of the galactic winds on the DGR evolution due to the dust growth in the ISM is opposite to that of the pre-enriched infall discussed above.

Figure~\ref{fig:DGR-O_wind} shows the predicted evolution of the DGR with O abundance for model galaxies with selective galactic winds. The differences with the reference models (Fig.~\ref{fig:DGR-O}) indeed become apparent with operation of the dust growth in the ISM, which happens at later times and lower metallicities in models with galactic winds. For example, in Models 1 and 7, the dust growth by accretion raises the DGR values by an order of magnitude at $\OH \sim 7$, which is 0.2~dex lower than in the reference models (Sect.~\ref{sec:DustEv}). Similar differences are present in other models with more intense star formation.

\begin{figure}[t]
\includegraphics[width=0.5\textwidth]{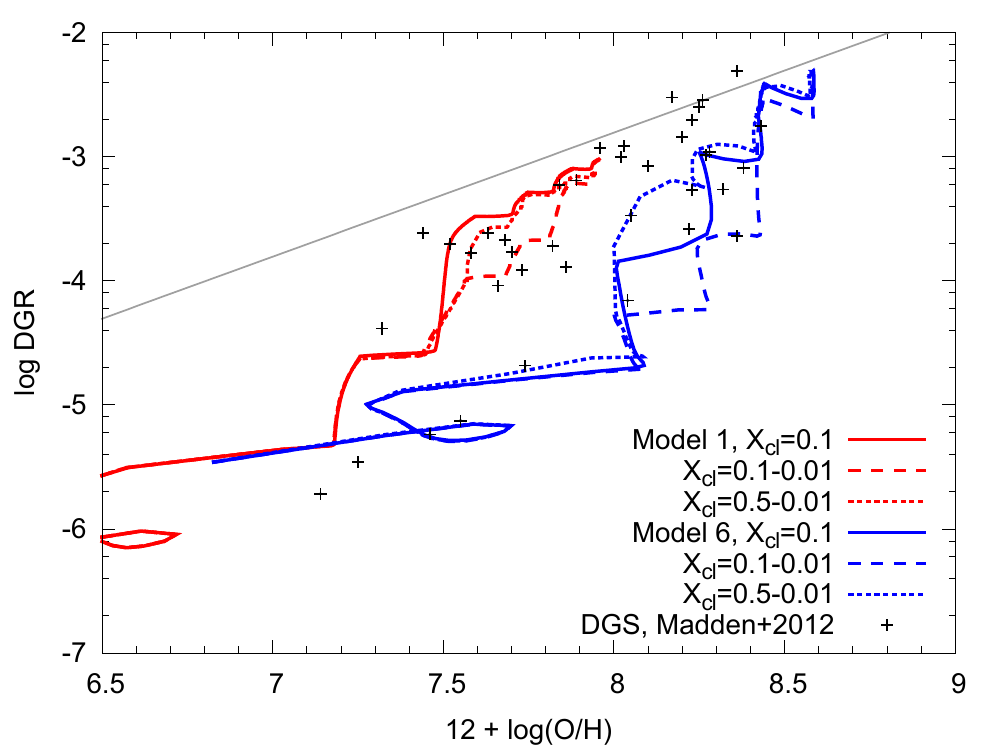}
\caption{Evolution of the dust-to-gas mass ratio with oxygen abundance for Models 1 and 6 with the fixed $X_{\rm cl} = 0.1$ (red and blue solid lines, respectively), and varying $X_{\rm cl}$ values during burst and quiescent epochs, of 0.1 and 0.01 (dashed lines of the same colors as the reference models) and 0.5 and 0.01 (dotted lines), respectively.}
\label{fig:DGR-O_Xcl}
\end{figure}

\subsubsection{Molecular gas content}\label{sec:CloudVar}
Dust production rate by the ISM growth in our model depends on the molecular gas mass fraction $X_{\rm cl}$  through the effective growth timescale given by Eq.~(\ref{Eq:EffTimesc}). Model calculations considered so far were performed with the fixed $X_{\rm cl}$ value of 0.1. 
We investigate how the variations in the dense gas fraction affect the dust growth in the ISM in model runs with the variable $X_{\rm cl}$ with the values of 0.1 and 0.5 during starbursts and a reduced cloud fraction $X_{\rm cl} = 0.01$ during quiescent phases. Figure~\ref{fig:DGR-O_Xcl} shows the DGR evolution with oxygen abundance for Model 1 with variable $X_{\rm cl}$ compared to the reference model (Table~\ref{tab:Models}). 

In calculations with $X_{\rm cl}$ value of 0.1 during starbursts, dependence on $X_{\rm cl}$ becomes apparent after the third burst of star formation, when the dust growth becomes an important dust source. A reduced cloud fraction during the post-burst evolution leads to a slower dust mass growth and lower DGR values compared to the reference model. However, the stepwise behaviour of the DGR indicates the importance of the dust growth during quiescent epochs, even when dense clouds constitute only 1\% by mass. When the dust growth process reaches saturation at $\ \gtrsim  7.8$, the DGR evolution does not depend on the dense cloud fraction, and both models predict similar DGR values. As expected, for a higher $X_{\rm cl}$ value of 0.5, saturation of the dust growth occurs earlier  (Fig.~\ref{fig:DGR-O_Xcl}). This is because the effective growth timescale during the fourth burst is comparable to a burst duration enabling a quick accretion of metals produced by massive stars formed in the same burst. In these calculations we neglect the delay introduced by cooling and mixing of hot supernova ejecta with dense gas before freshly synthesised metals can accrete onto grain surfaces. Hydrodynamical simulations of a gas rich dwarf galaxy with short starbursts showed that the mixing timescale of SN II ejecta with ambient gas can be very short, a few Myr \citep{Recchi:2001bp}. A more realistic value is probably a few hundred Myrs \citep[e.g.,][]{Yang:2012ia, Recchi:2013je}, which is shorter than the quiescent periods assumed here. 

For comparison, we also show the same calculations for Model 6, which has more frequent and more intense starbursts during the first Gyr of evolution (Fig.~\ref{fig:DGR-O_Xcl}). Dependence on the $X_{\rm cl}$ becomes evident before the third burst ($\OH\sim 8.1$) in the scatter of the DGR values. Afterwards, efficient dust growth by accretion takes place even during the starburst in the model with $X_{\rm cl}=0.5$. Similar to various runs for Model 1, the DGR values calculated with various $X_{\rm cl}$ for Model 6 converge towards the linear dust-to-metal scaling, but this happens  after the fourth starburst at evolution time $t \gtrsim 1$~Gyr and $\OH = 8.4$.
Thus, variations in the molecular gas fraction in dwarf galaxies introduce additional scatter in the DGR values and affect both the critical metallicity for the dust growth and the metallicity, at which dust growth reaches saturation.

\section{Conclusions}
\label{sec:Concl}
We studied the dust content of late-type dwarf galaxies with episodic star formation with  a multicomponent model of dust evolution. Our main result is that the growth of dust by accretion in the ISM can be important in dwarf galaxies, in contrast to early studies which considered only dust input from type II SNe. This leads to several consequences for dust evolution: (i)  the dust-to-gas ratio in young dwarf galaxies is determined by the dust input from stars characterised by much lower values than predicted by the linear dust-to-metal scaling;
(ii) with the ISM dust growth, the dust-to-gas ratio increases \textit{between starbursts}, while SNe II enrich the ISM with grains  during starbursts; 
(iii) critical metallicities for the dust growth in dwarfs are lower than in galaxies that continuously form stars because of long quiescent phases allowing accretion of existing metals after enrichment episodes.

The timescale of dust growth in the ISM is always limited by the accretion timescales, and thus by the amount of metals produced by a starburst. In metal-poor galaxies with slow enrichment, the dust growth in the ISM starts to dominate on a timescale of $\sim $1~Gyr, which is shorter than the ages of the old stellar populations found in many BCD galaxies. This timescale is comparable to the ages of AGB stars, therefore they are able to become the major dust sources in metal-poor galaxies with long quiescent epochs, before the growth in the ISM takes over dust production. 

Our models are able to reproduce the relation between the dust-to-gas ratio and oxygen abundance derived for a large dwarf galaxy sample from recent observations. A decrease in the DGR at low metallicities is explained by the dust input dominated by stars (SN~II and AGB stars) characterised by the relatively low efficiencies of dust condensation. 
Scatter in the dust-to-gas ratio observed at higher metallicities $7.2 < \rm log(O/H)+12<8.5$ stems from the fact that transition from the stardust- to the ISM-growth-dominated dust production happens at very different metallicities, depending on the star formation history. In galaxies with short bursts of star formation, accretion of metals on grains takes place mostly during quiescent phases of star formation, thereby steadily increasing the DGR for the same value of metallicity. Additionally, scatter in the dust-to-gas ratio is enhanced by the dust destruction during starbursts, resulting in temporal decreases in the dust-to-ratios. Variations in the dense cloud fraction expected in dwarf galaxies with episodic star formation also lead to a larger scatter in DGR--O relation. Galactic winds, which are commonly invoked to explain the low metallicities of dwarf galaxies, lead to the higher DGR values compared to the models without winds.

Calculations with increased efficiencies of dust condensation in type II SNe without dust growth by accretion in the ISM can reproduce neither the slope nor the scatter in the observed DGR versus metallicity relation. We conclude that the relation between dust content and metallicity in dwarf galaxies favours low condensation efficiencies in type II SNe, together with a gradual increase in the total dust mass by means of dust growth in the ISM.

\section*{Acknowledgements}
I acknowledge support by the \textit{Deutsche Forschungsgemeinschaft} through SPP 1573: ``Physics of the Interstellar Medium''. I thank Simone Recchi and Thomas Henning for stimulating discussions and careful reading of the manuscript.  I thank the GPU Cluster Milky Way at the FZ Juelich for computational resources. Finally, I am grateful to the referee, Gustavo Lanfranchi, for his useful comments and suggestions, which helped to improve the manuscript.

\bibliography{/Users/lana/Papers2/AllFromPapers}

\end{document}